\documentclass[prl,final,aps,showpacs]{revtex4}

\usepackage{graphicx}

\begin{document}


\title{Late time tails in the Kerr spacetime}
\author{Reinaldo J.~Gleiser}
\affiliation{Facultad de Matem\'atica, Astronom\'{\i}a y F\'{\i}sica,
Universidad Nacional de C\'ordoba,\\ Ciudad
Universitaria, 5000 C\'ordoba, Argentina}

\author{Richard H. Price}
\affiliation{Center for Gravitational Wave Astronomy and
Department of Physics and Astronomy, University of Texas at Brownsville,
Brownsville, Texas 78520
}

\author{Jorge Pullin} \affiliation{ Department of Physics and
Astronomy, Louisiana State University, Baton Rouge, LA 70803-4001}

\begin{abstract}
\begin{center}
{\bf Abstract}
\end{center}
Outside a black hole, perturbation fields die off in time as
$1/t^n$. For spherical holes $n=2\ell+3$ where $\ell$ is
the multipole index. In the nonspherical Kerr spacetime there is no
coordinate-independent meaning of "multipole," and a common sense
viewpoint is to set $\ell$ to the lowest radiatiable index, although
theoretical studies have led to very different claims. Numerical
results, to date, have been controversial.  Here we show that
expansion for small Kerr spin parameter $a$ leads to very definite
numerical results confirming previous theoretical predictions.
\end{abstract}

\maketitle

%
Perturbation fields outside a spherically symmetric black hole die off
in time in a way that has been well understood for more than 35
years\cite{1972PhRvD}.  This understanding is closely tied to
the fact that the spherical background allows the fields
to be decomposed into multipoles, each of which evolves from initial
data independently and can be studied with a relatively simple 1+1
computer code.  The evolved radiation field starts with oscillations
characteristic of the details of the initial data, then undergoes an
epoch of quasinormal ringing, and lastly falls off in time $t$ in the
form of a late-time ``tail'' $t^{-2\ell-3}$, where $\ell$ is the
multipole index\cite{moregeneral}.

\smallskip For the nonspherical Kerr black hole the situation has been
anything but clear since Kerr perturbations cannot be separated into
independently evolving multipoles\cite{notind}.  Certain symmetries do
apply, however. Perturbations can be separated into nonmixing
azimuthal Fourier modes $e^{im\phi}$ 
and into modes even or odd with respect to
reflection through the equatorial plane.  This has given
rise to what we shall call a ``common sense'' viewpoint in which
approximate spherical symmetry applies to the distant radiation field,
and the late-time behavior of the evolution of initial data is
dominated by the lowest multipole index (i.e.\,, the most slowly dying
tail) compatible with the azimuthal and equatorial symmetries of the
initial data. Thus, for example, a scalar perturbation field whose
initial data has $m=0 $ and that is symmetric with respect to the equator
will, at late time, be predominately a monopole and will have the
$t^{-3}$ tail of a monopole.

Supporting this viewpoint is the fact that, without spherical symmetry
in the background, there are no preferred angular coordinates. The
radial $r$ and polar $\theta$ coordinates are mixed differently in
different systems of coordinates used to describe the Kerr spacetime,
such as Boyer-Lindquist (BL) coordinates\cite{BL} or Kerr
coordinates\cite{Kcoords}. A ``multipole'' is specific to the
coordinate choice, and therefore cannot determine a physical effect,
like the rate of decrease of the field.

Theoretical work has argued against the common sense viewpoint, and
claims have appeared of numerical results to support both sides of the
argument.  The argument for something other than ``common sense'' was
first given by
Hod\cite{hod1998prd,hod1999hiorder,hod2000prdscalar,hod2000prl,hod2000prde+g},
(see also  Barack and Ori\cite{barackoriprl1999,barackprd2000}).  Hod
considered initial data that has only a single multipole $Y_{\ell m} $
in BL coordinates.  By looking at the zero frequency limit of a
Fourier transform, Hod argued that the tails of a massless scalar
function would have the following dependence on time and on
multipolarity of the initial data:
\begin{equation}\label{Hodrules} 
\Psi\propto\left\{
\begin{array}{ll}
Y_{\ell m}/t^{2\ell+3}\quad&\ell=m\  \mbox{ or}\ \ell=m+1\\
Y_{m m}/t^{\ell+m+1}\quad&\ell-m\geq2\ \ \mbox{(even)}\\
Y_{m\!+\!1\ m}/t^{\ell+m+2}\quad&\ell-m\geq2\ \ \mbox{(odd)}\,.
\end{array}
\right.
\end{equation}
More recently Poisson\cite{poissonprd2002} has come to the same
result, though with an approximate weak-field analysis.

The common sense results seemed so compelling, that numerical work was
immediately sought that would settle the issue, but numerical tests
required rather delicate 2+1 codes.  Krivan\cite{krivanprd1999} was
the first to attempt this work, using a scalar field with an initial
outgoing pulse with BL multipole indices $\ell,m=4,0$. The
Eq.~(\ref{Hodrules}) prediction for this case is a $t^{-5}$ monopole
tail while the common sense prediction is a $t^{-3}$ monopole. (This
$\ell,m=4,0$ case is the simplest scalar case for which there are
controversial predications, and we will consider it here as the
primary test case.)  Krivan's results weakly suggested a $t^{-5.5}$
law, but Krivan pointed out serious numerical problems caused by
angular differencing.  Subsequent computations of the $\ell,m=4,0$
case \cite{burkokhannaprd2003,scheeletalprd2004} found the common
sense $t^{-3}$ result, but those computations did not start with a
pure BL multipole on an initial hypersurface of constant BL time, so
that  neither of those studies represented the same problem as
that to which Eq.~(\ref{Hodrules}) and Krivan's results apply.

We present here a new approach for computationally probing the late
time evolution of tails in the Kerr spacetime and, in principle, in
other nonspherical spacetimes. The advantages of this approach are:
(i)~it gives a clear meaning to ``multipoles'' since it uses a
spherical operator for evolution; (ii)~there is no angular
differencing, and hence it avoids the errors  pointed out by
Krivan; (iii)~the
method gives convergent, clear numerical answers to the controversies
of tails in the Kerr spacetime.  The new approach expands fields and
the equations that govern them in powers of the spin parameter $a$, of
the Kerr metric.

We show here the approach as
applied to a scalar field $\psi$, both
for simplicity, and because previous numerical work has all been 
for a scalar field.  We make a further, minor simplification by
choosing the scalar field to have no $\phi$ (azimuthal) dependence.
More details and the case for more general fields will be published
elsewhere\cite{elsewhere}.

In Boyer-Lindquist coordinates, 
the Teukolsky equation for this case
(equivalent to $\psi_{,\alpha}^{\ ;\alpha}
=0$)  takes the explicit form
\begin{equation}\label{Teukeq} 
L[\psi]\equiv\left[\frac{
(r^2+a^2)^2}
{\Delta}-a^2\sin^2\theta
\right]\frac{\partial^2\psi}{\partial t^2}
-\Delta\frac{\partial^2\psi}{\partial r^2
}-2(r-M)\frac{\partial\psi}{\partial r}-\frac{1}{
\sin{\theta}}\,\frac{\partial}{\partial\theta}\left(
\sin\theta\frac{\partial\psi}{\partial\theta}
\right)=0
\end{equation}
where $\Delta$ has the usual meaning $r^2-2Mr+a^2$, with $M
$ the mass and $a=J/M
$ the angular momentum parameter.

For any initial data, the field $\psi $ that evolves will depend on the
spin parameter $a $ and we expand both $\psi $
and the operator of Eq.~(\ref{Teukeq}) in powers of $a/M$:
\begin{equation}
\psi=\psi^{(0)}+{a}/{M}\,\psi^{(1)}
+\left({a}/{M}\right)^2
\,\psi^{(2)}
+\cdots
\quad\quad L=L^{(0)}+\left({a}/{M}\right)^2L^{(2)}+\cdots
\end{equation}
Here 
the $\psi^{(n)}$ are functions of $t,r$ and $\theta$.
The equations that result 
for the even powers of $a/M
$ are
\begin{eqnarray}
L^{(0)}[\psi^{(0)}]&=&0\label{cascade1}  \\
L^{(0)}[\psi^{(2)}]&=&-L^{(2)}[\psi^{(0)}]\label{cascade2}\\
L^{(0)}[\psi^{(4)}]&=&-L^{(4)}[\psi^{(0)}]-L^{(2)}[\psi^{(2)}]\label{cascade3}\\
L^{(0)}[\psi^{(6)}]&=&-L^{(6)}[\psi^{(0)}]-L^{(4)}[\psi^{(2)}]-L^{(2)}[\psi^{(4)}]
\label{cascade4}
\end{eqnarray}
and so forth. 
The evolved fields can  be solved order by order.  Since the
right hand sides in the above sequence are treated as known driving
terms, only the operator $L^{(0)}$ need be inverted, but $L^{(0)}$
is just the spherically symmetric Schwarzchild operator, so the
multipoles contained in the solutions will only be those that appear
in the driving terms.

We focus now on the primary test case $\ell=4$, and on the question of
the exponent for the late time tail.  Since the initial data has
$\ell=4 $ there will be an evolved field that is zero order in $a/M$,
i.e., the purely $\ell=4$ field that is evolved by the Schwarzschild
operator $L^{(0)}$ in Eq.~(\ref{cascade1}).  The $\ell=4$ zeroth order
field $\psi^{(0)}$ will provide a source term on the right hand side
of Eq.~(\ref{cascade2}). But $L^{(2)}$ (the only term in the full set
of $L^{(n)} $ that is not spherically symmetric) contains a term
$-M^2\sin^2\theta\partial^2_t$, and the $\sin^2\theta$ multiplied by
$P_\ell(\cos\theta)$ gives multipoles of order $\ell-2,\; \ell$, and
$\ell+2 $. Thus $\psi^{(2)}$ in Eq.~(\ref{cascade2}) will be driven by
source terms with $\ell=$ 2,4, and 6. This changing of the multipole
order by $L^{(2)}$ will happen again in Eqs.~(\ref{cascade3}) and
Eq.~(\ref{cascade4}), since they also contain $L^{(2)} $.  As a result
of this multipole mixing, initial zero-order $\ell=4 $ data results in
a monopole field with terms of order 4,6,8,\ldots.  The 
sequence of multipole couplings from the zeroth order $\ell=4 $ to the 
final fourth order monopole can happen in only one way:
\begin{equation}
\psi^{(0)}_{\ell=4}\rightarrow \psi^{(2)}_{\ell=2}\rightarrow 
\psi^{(4)}_{\ell=0}\,.
\end{equation}
The sixth order monopole, however, has contributions from the zeroth order $\ell=4
$ field along 3 different paths of coupling:
\begin{equation}
\psi^{(0)}_{\ell=4}\rightarrow \psi^{(2)}_{\ell=4}\rightarrow 
\psi^{(4)}_{\ell=2}\rightarrow\psi^{(6)}_{\ell=0}\quad\quad
\psi^{(0)}_{\ell=4}\rightarrow \psi^{(2)}_{\ell=2}\rightarrow 
\psi^{(4)}_{\ell=2}\rightarrow\psi^{(6)}_{\ell=0}\quad\quad
\psi^{(0)}_{\ell=4}\rightarrow \psi^{(2)}_{\ell=2}\rightarrow 
\psi^{(4)}_{\ell=0}\rightarrow\psi^{(6)}_{\ell=0}\quad\quad
\end{equation}

The four wave equations (\ref{cascade1})--(\ref{cascade4}) can now be
decomposed into multipoles, resulting in six 1+1 wave equations involving
radius and time.  There is, however, a technical complication: 
singularities
appear in the driving terms at $r=2M$. This coordinate effect, due 
to the fact that the $r
$ coordinate location of  Kerr horizon 
depends on $a$, can be removed by introducing a new 
radial coordinate $\rho
$
\begin{equation}
r=M+\sqrt{\rho^2-2\rho M+M^2-a^2}\,,
\end{equation}
so that the Kerr horizon is at $\rho=2M
$ independent of $a
$. Along with this new radius, we use its associated ``tortoise'' version 
 $\rho^*=\rho+2M\log{(\rho-2M)}$.

The 1+1 wave equations are most conveniently written if we introduce the notation
$\psi^{(\rm order)}(t,\rho,\theta)={\rho}^{-1}
\sum_\ell f^{(\rm order)}_\ell
(t,\rho)P_{\ell}(\cos\theta)$, for multipole decompositions.
The six equations we need then are
\begin{eqnarray}
\partial^2_tf^{(0)}_{4}-\partial^2_{\rho*}f^{(0)}_{4}
+\frac{1-2M/\rho}{\rho^2}\left(20+\frac{2M}{\rho}\right)f^{(0)}_{4}&=&0\label{feq1}
\\
\partial^2_tf^{(2)}_{2}-\partial^2_{\rho*}f^{(2)}_{2}
+\frac{1-2M/\rho}{\rho^2}\left(6+\frac{2M}{\rho}\right)
f^{(2)}_{2}&=&-\frac{4}{21}\,
\frac{M^2}{\rho^2}\,\left(1-\frac{2M}{\rho}\right)\partial^2_tf^{(0)}_{4} 
\label{feq2}\\
\partial^2_tf^{(2)}_{4}-\partial^2_{\rho*}f^{(2)}_{4}
+\frac{1-2M/\rho}{\rho^2}\left(20+\frac{2M}{\rho}\right)
f^{(2)}_{4}&=&\frac{2}{77}\frac{M^2(19+20M/\rho+38M^2/\rho^2
)
}{\rho^2(1-M/\rho)}\,\partial^2_tf^{(0)}_{4}\nonumber\\
-\frac{M^2}{\rho^2(1-M/\rho)^2}\partial^2_{\rho*}f^{(0)}_{4}
+\frac{M^2(1-2M/\rho}{\rho^3(1-M/\rho)^3}\partial_{\rho*}f^{(0)}_{4}
&-&\frac{M^2(1-2M/\rho)(1-4M/\rho+2M^2/\rho^2)}{\rho^4(1-M/\rho)^3}
f^{(0)}_{4}\label{feq3}\\
\partial^2_tf^{(4)}_{0}-\partial^2_{\rho*}f^{(4)}_{0}
+\frac{1-2M/\rho}{\rho^2}\left(\frac{2M}{\rho}\right)f^{(4)}_{0}&=&-\frac{2}{15}\,
\frac{M^2}{\rho^2}\,\left(1-\frac{2M}{\rho}\right)\partial^2_tf^{(2)}_{2}\,.\label{feq4}\\
\partial^2_tf^{(4)}_{2}-\partial^2_{\rho*}f^{(4)}_{2}
+\frac{1-2M/\rho}{\rho^2}\left(6+\frac{2M}{\rho}\right)f^{(4)}_{2}&=&
\frac{2}{21}\,\frac{M^2(5+6M/\rho+10M^2/\rho^2)}{\rho^2(1-M/\rho)}
\partial^2_tf^{(2)}_{2}
-\frac{M^2}{\rho^2(1-M/\rho)^2}\,\partial^2_{\rho*}f^{(2)}_{2}\nonumber\\
-\frac{M^2(1-2M/\rho)(1-4M/\rho+2M^2/\rho^2)}{\rho^4(1-M/\rho)^3}f^{(2)}_{2}
&+&\frac{M^2(1-2M/\rho)}{\rho^3(1-M/\rho)^3}\partial_{\rho*}f^{(2)}_{2}
-\frac{4}{21}\frac{M^2(1-2M/\rho)}{\rho^2}\partial^2_{t}f^{(2)}_{4}
\,.\label{feq5}\\
\partial^2_tf^{(6)}_{0}-\partial^2_{\rho*}f^{(6)}_{0}
+\frac{1-2M/\rho}{\rho^2}\left(\frac{2M}{\rho}\right)f^{(6)}_{0}&=&
\frac{2}{3}\,\frac{M^2(1+2M^2/\rho^2)}{\rho^2(1-M/\rho)}
\partial^2_{t}f^{(4)}_{0}-\frac{M^2
}{\rho^2(1-M/\rho)^2}\partial^2_{\rho*}f^{(4)}_{0} \nonumber\\
+\frac{M^2(1-2M/\rho)}{\rho^3(1-M/\rho)^3}\partial_{\rho*}f^{(4)}_{0}
&-&\frac{2}{15}\frac{M^2(1-2M/\rho) }{\rho^2}\partial^2_{t}f^{(4)}_{2}\label{feq6}\,.
\end{eqnarray}
Numerical computations with these equations were carried out on a $t,\rho*$
characteristic grid, with no boundary conditions. (Computations were
carried out only in the domain of dependence of the initial spatial
grid.) All six fields $f^{(0)}_{4}, f^{(2)}_{4}, f^{(2)}_{2},
f^{(4)}_{2},f^{(4)}_{0}, f^{(6)}_{0}, $ were evolved simultaneously.
Initial data for $f^{(0)}_{4}$, at $t=0$, was chosen to be a Gaussian
pulse and was made (approximately) outgoing by taking the initial
value at grid point $\rho^* $ to be replicated after a time step
$\Delta t$, at the spatial grid point $\rho^*+\Delta t$. Initial data
were taken to be zero for all higher order fields, so that the final
monopole was only the result of the initial pure $\ell=4 $
data. Except for the final monopole, the resulting late-time
power-laws, at every order, followed the common-sense 
$2\ell+3$ rule. The zeroth-order 
$\ell=4$ field, for example, falls off as $1/t^{11}$; the second-
and fourth-order quadrupole fields fall off as 
$1/t^{7}$. 

\begin{figure}[ht]
\includegraphics[width=.35\textwidth]{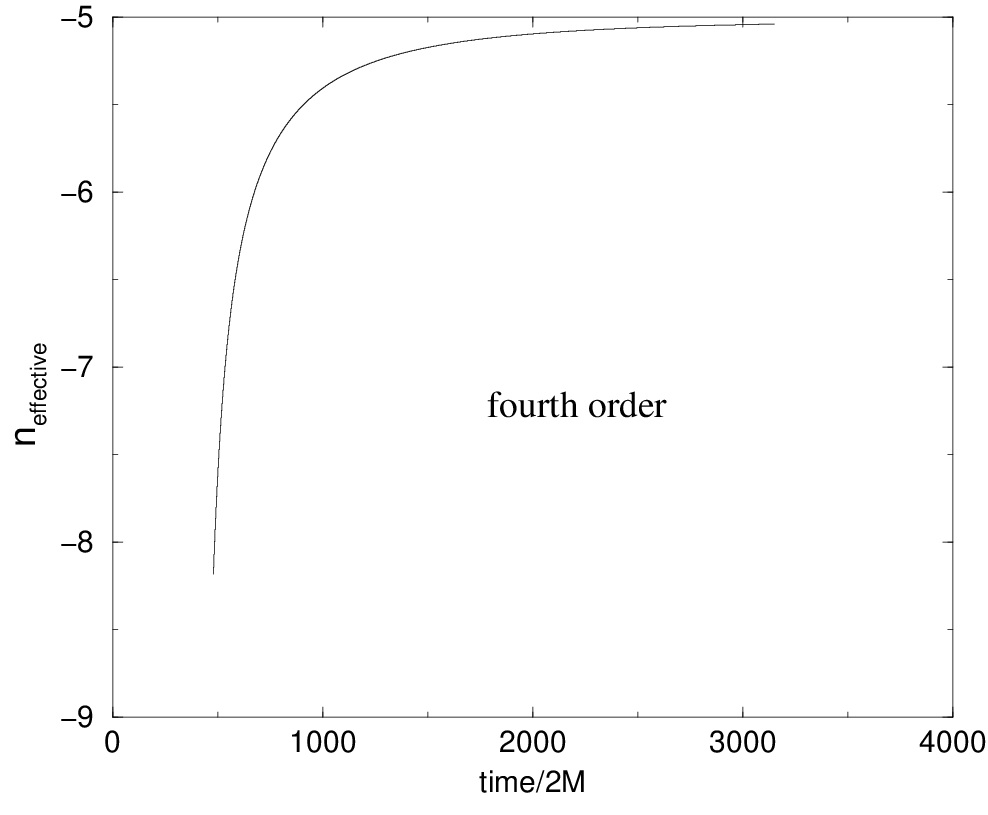}\hspace{10pt}
\includegraphics[width=.35\textwidth]{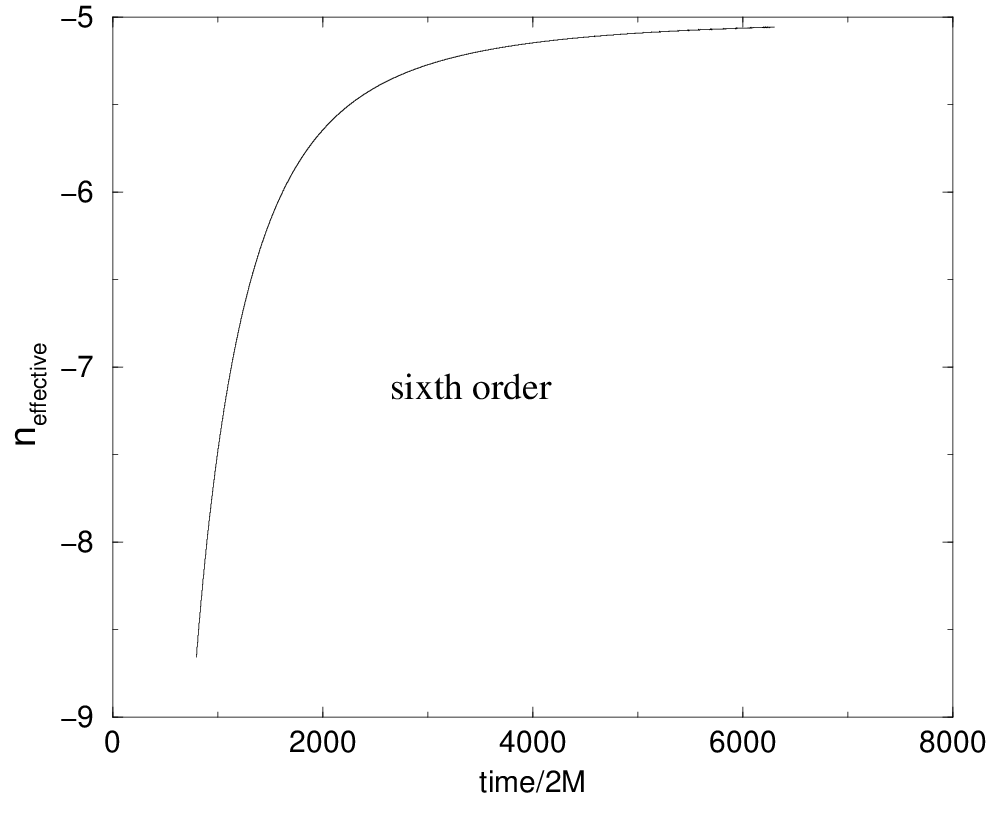} \caption{The effective
power-law index $d\log{f}/dt$ for both the fourth order and sixth
order terms in the monopole moment. Initial data in both cases was an
outgoing pulse with initial profile $\exp{[-(\rho*+100M)^2 /(40M)^2 ]
} $ } \label{fig:neffective}
\end{figure}

Only the monopole gives rule-breaking results.
Results are shown in Fig.~\ref{fig:neffective} for the effective
power-law index $n_{\rm effective} =d\log{f}/d\log{t}$ of both
$f^{(4)}_{0}$ and
$f^{(6)}_{0} $, the fourth- and sixth-order monopoles.
The cascade of equations (\ref{feq1})--(\ref{feq6}) links the highest
order and lowest order fields, in effect, by high-order derivatives,
so the results shown in Fig.~\ref{fig:neffective} required smoothing
of the computational output as will be described in a longer
paper\cite{elsewhere}. 
The results in Fig.~\ref{fig:neffective} show clearly that the late
time tail is characterized by a $1/t^5$ fall off, not the more
``sensible'' $1/t^3 $ fall off. The default grid size used was $\Delta
t=\Delta\rho*=0.09M $, but the results for $n_{\rm effective} $ were
unchanged for a reasonable variation in the grid size. 

From a numerical point of view the $1/t^5 $ result is truly
remarkable. The 
$\ell>0
$ fields die off very quickly, so it should be valid to consider 
the late-time monopole evolving without source.
The monopole {\em should} have the exponent $n=3.$ Indeed, small
modifications in the computation do change the exponent from $n=5$ to
$n=3$. We see this change if we put in nonzero initial data for
$f^{(4)}_{0}$, or for $f^{(2)}_{2}$. We see this change also if we
arbitrarily turn off the evolution of the $f^{(0)}_{4}$ field at some
intermediate time and let the higher order  fields
continue to evolve.  The overwhelming tendency for the exponent to be
3 rather than 5 convinces us that there is no error we have overlooked
in our program; any error would almost surely lead to $n=3$.

The delicacy of the the $n=5$ result underscores the numerical
advantages of the 1+1 computations in Eqs.~(\ref{feq1}) --
(\ref{feq6}) over 2+1 codes, even though the set of equations contains
second derivatives applied several times on the right hand source
terms. The method turns out to be accurate enough that we have been
able to go one step further, and treat the case of initial
$\ell,m=6,0$ data (as yet only to fourth order) for which the common
sense monopole exponent is $n=3 $, while Eqs.~(\ref{Hodrules})
predicts $n=7$. We have found, with an accuracy equivalent to that
shown in Fig.~\ref{fig:neffective}, that the late-time monopole has
index $n=7$.

In a forthcoming paper\cite{elsewhere} we shall present further
details of the method used and shall also provide a wider set of
numerical examples, including those for nonaxisymmetric initial data
and for gravitational perturbations, and tests of Poisson's
approximation\cite{poissonprd2002}.

During a revision of this paper we learned from Gaurav
Khanna\cite{khannaprivate} that he had been able to evolve the scalar
Teukolsky equation with a 2+1 code from pure BL initial data.  A
comparison of results, for small and moderate $a/M $ showed excellent
agreement with our results.

We gratefully acknowledge support by the National Science Foundation
under grants PHY0554367 (RHP) and PHY0545837 (JP). The work of RJG is
supported by CONICET (Argentina) and by grants from CONICET and
Universidad Nacional de C\'ordoba.  We also acknowledge the support of
the NASA Center for Gravitational Wave Astronomy at University of
Texas at Brownsville, the Horace Hearne Jr.\ Institute for Theoretical
Physics, and CCT-LSU.  We thank Gaurav Khanna for sharing with us 
his numerical results.
For useful suggestions and discussions of this
problem we thank Eric Poisson, Leor Barack, Lior Burko, 
William Krivan, Luis Lehner, Amos Ori, Manuel Tiglio, and the members of the 
relativity group at the University of Wisconsin at Milwaukee.


\begin{thebibliography}{10}

\bibitem{1972PhRvD}
R.~H. {Price}, \prd {\bf 5},  2419  (1972).

\bibitem{moregeneral}
This index applies to the case that the initial data is compact and has a
  nonzero initial time derivative. For simplicity we will consider only this
  case.

\bibitem{notind}
In the Fourier domain the Teukolsky equation can be separated into multipoles,
  but the separation constants are functions of frequency. In the time domain,
  therefore, the multipoles interact.

\bibitem{BL}
R.~H. {Boyer} and R.~W. {Lindquist}, J. Math. Phys. {\bf 8},  265  (1967).

\bibitem{Kcoords}
R.~P. {Kerr}, Physical Review Letters {\bf 11},  237  (1963).

\bibitem{hod1998prd}
S. {Hod}, \prd {\bf 58},  104022  (1998).

\bibitem{hod1999hiorder}
S. {Hod}, \prd {\bf 60},  104053  (1999).

\bibitem{hod2000prdscalar}
S. {Hod}, \prd {\bf 61},  024033  (2000).

\bibitem{hod2000prl}
S. {Hod}, Physical Review Letters {\bf 84},  10  (2000).

\bibitem{hod2000prde+g}
S. {Hod}, \prd {\bf 61},  064018  (2000).

\bibitem{barackoriprl1999}
L. {Barack} and A. {Ori}, Physical Review Letters {\bf 82},  4388  (1999).

\bibitem{barackprd2000}
L. {Barack}, \prd {\bf 61},  024026  (2000).

\bibitem{poissonprd2002}
E. {Poisson}, \prd {\bf 66},  044008  (2002).


\bibitem{krivanprd1999}
W. {Krivan}, \prd {\bf 60},  101501  (1999).

\bibitem{burkokhannaprd2003}
L.~M. {Burko} and G. {Khanna}, \prd {\bf 67},  081502  (2003).

\bibitem{scheeletalprd2004}
M.~A. {Scheel} {\it et~al.}, \prd {\bf 69},  104006  (2004).


\bibitem{elsewhere}
R.~J. {Gleiser}, R.~H. {Price}, and J. {Pullin}, paper in preparation.

\bibitem{khannaprivate}
Private communication.
\end{thebibliography}
\end{document}